\documentclass[aps,prl,12pt,twocolumn,superscriptaddress,showpacs,floatfix,reprint]{revtex4-1}

\usepackage{amsmath,amssymb}
\usepackage{graphicx,float}
\usepackage{bm}
\usepackage{mathrsfs}
\usepackage{rotating}
\usepackage[amssymb]{SIunits}
\usepackage{inputenc}

\usepackage{color}

\allowdisplaybreaks[2]

\begin{document}

\title{Solid state optical interconnect between distant superconducting quantum chips}

 \author{Keyu \surname{Xia}}  %
  \email{keyu.xia@mq.edu.au}
 \affiliation{ARC Centre for Engineered Quantum Systems, Department of Physics and Astronomy, Macquarie University, NSW 2109, Australia}

 \author{Jason \surname{Twamley}}
  \email{jason.twamley@mq.edu.au}
 \affiliation{ARC Centre for Engineered Quantum Systems, Department of Physics and Astronomy, Macquarie University, NSW 2109, Australia}

\begin{abstract}
%
We propose a design for a quantum interface exploiting the electron spins in crystals to swap the quantum states between the optical and microwave. Using  sideband driving of a superconducting flux qubit and a combined cavity/solid-state spin ensemble Raman transition,  we demonstrate how a stimulated Raman adiabatic passage (STIRAP)-type operation can swap the quantum state between a superconducting flux qubit and an optical cavity mode with a fidelity higher than $90\%$. We further consider two distant superconducting qubits with their respective interfaces joined by an optical fiber and  show a quantum transfer fidelity exceeding $90\%$ between the two  distant qubits.
\end{abstract}


\maketitle
Superconducting qubits (SQs) are one of the most promising technologies for delivering a  quantum computer which will operate within a single superconducting chip \cite{SQ1}. Linking remotely distant superconducting chips via an optical data bus thus opens the door for a quantum internet \cite{QNet1,QNet2,QNet3,Hammerer:2010gs, Ritter:2012jn}.
To permit the optical  networking of superconducting quantum chips \cite{SQ1}, a coherent quantum interconnect between microwave and optical quantum information as a building block has been developed theoretically and experimentally \cite{Stannigel:2010eu, Tian:2012jb,Wang:2012em, Barzanjeh:2012ez, Xiang:2013hm,Hafezi:2012gz,KXMVJT:2014SRep}.
 One  way to achieve this interlinking is to find a method to convert  quantum information held within superconducting circuits into quantum information held in optical photons within one node (or interface), then transmit the photons over an optical fibre, and finally reconvert the photonic quantum information back into the target distant superconducting chip. 

Researchers have recently focused much effort towards devising  such a quantum interface, interconverting electromagnetic radiation at different frequencies using either optomechanical quantum systems \cite{Stannigel:2010eu,  Tian:2012jb,Wang:2012em, Barzanjeh:2012ez, Xiang:2013hm,KXMVJT:2014SRep}, or ensemble of ultracold atoms \cite{Hafezi:2012gz}. Although solid-state electronic ensembles have been experimentally demonstrated to be capable of coherent exchange with superconducting circuits \cite{Kubo:2010iq, Schuster:2010bu, Bushev:2011be, Zhu:2012js}, only two proposals for a solid-state hybrid quantum interface have been proposed very recently \cite{MWOptConv3,MWOptConv4}, and one more abstract scheme for  quantum state transfer between two remote NV centers \cite{Stannigel:2010eu}. Experimental progress in  magneto-optic frequency conversion exploiting  optomechanical resonators has been demonstrated  for  classical signals \cite{MWOptConv1,MWOptConv2}. We have recently described a design for a magneto-optomechanical quantum interface for and detailed how it can achieve high transfer fidelities between distant flux qubits \cite{KXMVJT:2014SRep}. All of the published proposals so far are technically challenging, using either optomechanical/magneto-optomechanical systems, or ensembles of trapped atoms in proximity to cryogenic superconducting circuits except \cite{MWOptConv4}. 

In this Letter we show how such an interconversion between microwave and optical quantum information and a quantum internet can be achieved through a solid-state electronic spin ensemble which catalyses the coupling between the flux qubit and photonic cavity mode. 

\begin{figure}
\centering
\setlength{\unitlength}{1cm}
\includegraphics[width=0.6\linewidth]{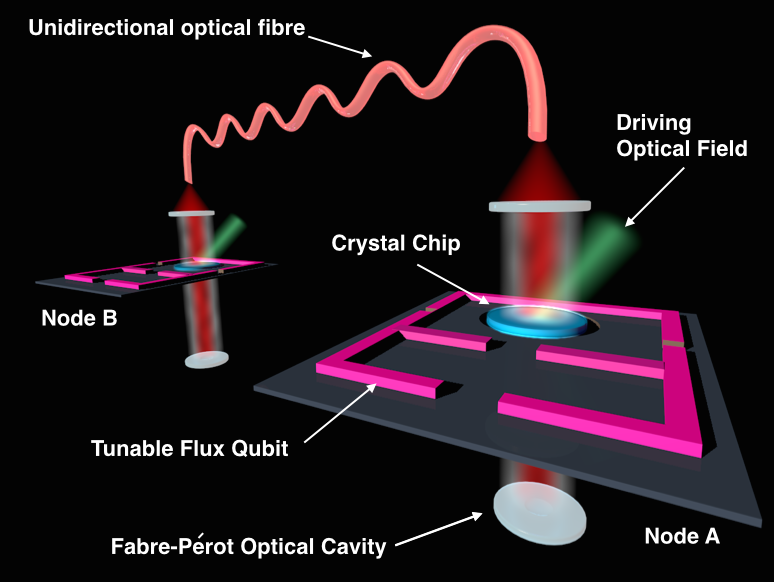}\\
\includegraphics[width=0.6\linewidth]{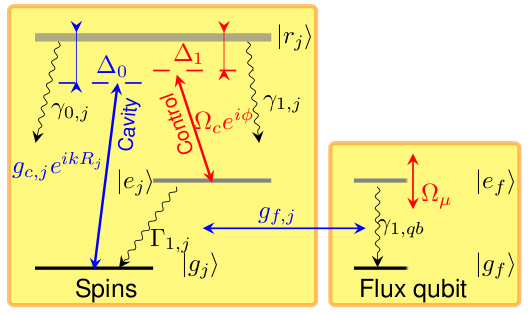}\\
\caption{(Color online) Schematic design and operation of the quantum network. (a) Schematic of quantum network between two superconducting flux qubits. Node A and B are identical and connected by an one-way optical fiber. Each node consists of a four-junction flux qubit and an ensemble of electronic spins.  The flux qubit is tuned by flux biases $\Phi_b$ and $\Phi_\alpha(t)$. In each node, the spins interact  magnetically with a flux qubit and optically with a cavity mode $\hat{a}_c$ and a coherent optical driving field $\Omega_c$.
(b) Level diagram describing the interactions between the $j$th spin and the flux qubit and the optical cavity mode. 
Each spin is modeled as a $\Lambda$-type three-level system, while the gap tunable flux qubit is modeled as a two-level system. The coherent (cavity) optical field drives the transition $|r_j\rangle \rightarrow |e_j\rangle (|g_j\rangle)$. The flux qubit magnetically drives the transition $|e_j\rangle \rightarrow |g_j\rangle$.
}\label{fig:setup}
\end{figure}


We begin by describing the schematic configuration for our quantum interface/node and later how to use two distant nodes to achieve quantum state transfer i.e. a quantum network  (Fig.~\ref{fig:setup}). Each node works as a quantum interface and composes of a Fabry-P\'erot cavity, an ensemble of spins and a gap tunable four-Josephson-junction (4JJ) flux qubit. 
This 4JJ flux qubit is fabricated on top of a suitable substrate (dark plane) Fig.~\ref{fig:setup}. In order to optically couple the ensemble of spins in a crystal chip, a round hole is cut in the main loop of the flux qubit such that the light in the Fabry-P\'erot cavity can pass through it without significant absorption and scattering. The crystal chip covers the hole. The crystal chip is cut  so that the [001] direction is along the $z-$axis, perpendicular to the plane of the flux qubit and parallel with the axis of the optical cavity. In this arrangement, the spin ensemble can effectively couple to both of the cavity mode and the magnetic field created by the flux qubit (see Supplementary Material for more details). We consider that an optical coherent control field irradiates the crystal chip at almost normal incidence to the chip.
We first discuss the swapping of quantum information between the flux qubit and the optical cavity mode in a single node/interface but later will consider 
the quantum transfer between two nodes $A$ (the source), and $B$ (the target), which are optically connected via a unidirectional optical fiber. The latter only allows optical propagation along one direction, from  node $A$ to  node $B$ and can be realized using  optical circulators \cite{1997PhRvL..78.3221C}.

We now describe the optical subsystem of each node and the unidirectional optical fibre couplings. We assume that the optical cavity mode $\hat{a}_{c,l}$ in the $l^{th}$ node ($l \in \{A,B\}$) has a resonance frequency $\omega_{c,l}$ and an intrinsic decay rate $\kappa_i^{(l)}$. It couples to an output optical fiber with a coupling  rate $\kappa_{ex}^{(l)}$ and connects to the other node via this fiber. Thus the total decay rate becomes $\kappa_l=\kappa_i^{(l)} + \kappa_{ex}^{(l)}$, and we assume $\kappa_{ex}^{(l)} = \xi_l \kappa_l$ with $0 \leqslant \xi_l\leqslant 1$. $\xi_l<1$ represents the photon loss in the transfer. The  $\kappa_{ex}^{(l)}$ denote the couplings into the waveguide connecting the two distant nodes. In this two node network, the Lindblad superoperator describing the one-way quantum connection breaking the time symmetry is given by
$ \hat{\hat{L}}_{Net} = -\sqrt{\kappa_{ex,A}\kappa_{ex,B}}\left(\hat{a}_B^\dag\hat{a}_A\rho - \hat{a}_A\rho\hat{a}_B^\dag + \rho\hat{a}_A^\dag\hat{a}_B - \hat{a}_B\rho\hat{a}_A^\dag  \right)$
, where $\hat{a}_A^\dag$ ($\hat{a}_B^\dag$) and $\hat{a}_A$ ($\hat{a}_B$) are the creation and annihilation operators of the cavity mode in node A (B).

We next describe the crystal chip spin ensemble. We model the collection of electron spins as ensemble of three-level systems, as shown in Fig. \ref{fig:setup}(b). These could be either a collection of $\text{NV}^-$ centers in diamond \cite{EnsembleNVEIT1} or rare earth ion crystal such as $\text{Er}^{3+}$ ions in  $\text{Y}_2\text{SiO}_5$ \cite{EnsembleErEIT3,EnsembleErEIT5}. Each spin system consists of three levels: the optical excited state $|r_j\rangle$, and two electronic ground states  $|g_j\rangle$ and $|e_j\rangle $.  In the case of the $\text{NV}^-$ we have 
$\{ |r_j\rangle, 
|g_j\rangle,
|e_j\rangle\}=\{
 |^3{\bf E},S_z\rangle_j,
|{}^3{\bf A},m_s=0\rangle_j, 
(|{}^3{\bf A},m_s=+1\rangle_j + |{}^3{\bf A},m_s=-1\rangle_j)/\sqrt{2}\}$, while for the $\text{Er}^{3+}$ system we have $\{ |r_j\rangle, |g_j\rangle,|e_j\rangle\}=\{|^4{\bf I}_{13/2},Y_1\rangle_j, 
|^4{\bf I}_{15/2},Z_1, m_s =-1/2\rangle_j, |^4{\bf I}_{15/2},Z_1, m_s = +1/2\rangle_j\}$.
 The excited state $|r_j\rangle$ decays to the state $|g_j\rangle$ ($|e_j\rangle$) with a rate $\gamma_{0,j}$ ($\gamma_{1,j}$). Due to the coupling to the magnetic environment, the state $|e_j\rangle$ also decays to $|g_j\rangle$ with the rate $\Gamma_{1,j}$ and suffers pure dephasing of a rate $\Gamma_{\phi,j}$. We can describe the ensemble of spins within each  each node with the Hamiltonian ($h=1$) $H_{Spin}= \sum_{j}\frac{D_j}{2}S_{z,j} + g_e\mu_B {\bf B}_{z,j} \cdot {\bf S}_{z,j} + \omega_{r,j}|r\rangle_j \langle r_j| $, where $S_{z,j}$ is the $z$-component of the usual spin-1/2 Pauli operators ${\bf S}_j$ for the $j$th spin, $D_j$ is the zero-field splitting with $D_j \approx 2.8$ \giga\hertz[$4$ \giga\hertz~] for the $\text{NV}^-$[$\text{Er}^{3+}$] centers, $\omega_{r,j}$ is the energy of the optical excited state $|r_j\rangle$. The second term in the above Hamiltonian describes the magnetic interaction with the spins with $g_e$ being the Land\'e $g$ factor of the spin. Note that the $g$-factor of the electronic spin in $\text{Er}^{3+}$  can be up to $g_e\sim 15$ \cite{ErMagTran1}, which is  much larger than in $\text{NV}^-$ centers where $g_e\sim 2$ and $\mu_B = 14$ \mega\hertz\cdot\milli\tesla$^{-1}$ is the Bohr magneton. Here we neglect the terms related to the strain-induced splitting because they are very small and only shift the energy of state $|e_j\rangle$.

 The 4JJ flux qubit can be modeled as a two-level system with the excite state $|e_f\rangle$ and the ground state $|g_f\rangle$, and can be tuned by the flux biases $\Phi_{b}$ and $\Phi_{\alpha}$. The flux $\Phi_{\alpha}$ threading through the $\alpha$ loop is used to tune the gap $\mathscr{T}(\Phi_{\alpha})$. We apply a time-dependent magnetic field $\Phi_\mu(t) = A_\mu(t) \cos(\omega_\mu t)$ to the $\alpha$ loop and thus tune the gap of qubit. Thus the free Hamiltonian of the gap tunable flux qubit becomes $H_Q=\omega_q/2 \sigma_z + \Omega_\mu \cos(\omega_\mu t) \sigma_z$, where $\omega_q = \sqrt{\mathscr{T}^2 + \in^2}$ with $\in(\Phi_b) = 2I_p (\Phi_b -\Phi_0/2)$,  ($\Phi_b$ is the external flux threading the qubit loop, $\Phi_0$ the flux quantum and $I_p$ the persistent current), is the energy bias and $\mathscr{T}$ is the tunnel splitting dependent on the  bias $\Phi_{\alpha}$, and $\Omega_\mu \varpropto A_\mu$. In the vicinity of $\Phi_b \approx \Phi_0/2$ we have $\mathscr{T}(\Phi_\alpha) \gg \in$. The flux qubit associated Pauli spin-half operators are defined as $\sigma_z = |e_f\rangle \langle e_f| - |g_f\rangle \langle g_f|$ and $\sigma_x = |e_f\rangle \langle g_f| + |g_f\rangle \langle e_f|$. The decay and pure dephasing rates of the excited state of the flux qubit $|e_f\rangle$ are denoted as $\gamma_{1,qb}$ and $\gamma_{2,qb}^*$, respectively.

We now describe the tripartite interaction between the ensemble of spins, the cavity mode and the flux qubit within an individual node and which are graphically summarized in Fig \ref{fig:setup}(b). We consider an optical Raman transition between the two spin ground states $|g_j\rangle$ and $|e_j\rangle$ formed through a combination of the external driving classical optical field and the  quantum  optical cavity field. This optical driving of an individual spin at  position ${\bf R}_j$ together with the cavity mode $\hat{a}_{c,j}$ drives the transition of $|g_j\rangle \leftrightarrow |r_j\rangle$ with a rate $g_{c,j}$ and a phase $\theta_j={\bf k\cdot R}_j$, where the latter is dependent on the position ${\bf R}_j$ and the wave vector ${\bf k}$ of $\hat{a}_{c,j}$. The the coherent laser field $\Omega_{c,j}$, with the frequency $\omega_L$ and  chirped phase $\phi(t)$, drives the transition of $|e_j\rangle \leftrightarrow |r_j\rangle$. The fluctuation in the coherent driving is taken into account by $\theta_j$ and $g_{c,j}$. In the dispersive regime, these optical transitions form a Raman transition between the two ground states $|g_j\rangle$ and $|e_j\rangle$, which is also arranged to dispersively couple to the flux qubit. Such optical $\Lambda-$type configuration has been demonstrated in both  ensemble of $\text{NV}^-$ centers \cite{EnsembleNVEIT1}, and $\text{Er}^{3+}$ host crystals \cite{EnsembleErEIT1,EnsembleErEIT2,EnsembleErEIT3}. The diamond crystal chip is synthesized with (001) surface orientation such that the magnetic field generated by the flux qubit has a component orthogonal to the principal ($z$) axes of the NV centers \cite{Zhu:2012js}. The total Hamiltonian of the coupled system within one node is 
$H =  \frac{\omega_q}{2} \sigma_z + \Omega_\mu \cos(\omega_\mu t) \sigma_z + \sum_j^N g_{f,j}\sigma_x S_{x,j}
       + \sum_j^N \frac{D_j}{2} S_{z,j} + \omega_{r,j}|r_j\rangle \langle r_j| + \omega_c \hat{a}^\dag \hat{a} 
       +\sum_j^N \left( g_{c,j}e^{ikR_j} \hat{a}^\dag |g_j\rangle \langle r_j| 
       +  \Omega_c  e^{i\omega_L t + i\phi(t)} |e_j\rangle \langle r_j| + H.c. \right)$.
%

We consider the first-order sideband transition and adiabatically eliminate the optical excited state $|r_j\rangle$, then the Hamiltonian in the one-excitation space (OES) becomes (see supplementary information)
\begin{equation} \label{eq:H}
\begin{split}
  H & =  (\dot{\phi} + \tilde\Delta_q) \sigma_+ \sigma_- -\sum_j^N  \Delta_{j} \bar{S}_{+,j}\bar{S}_{-,j} - \delta_{en} \hat{a}^\dag \hat{a}\,\\
      & - \sum_j^N \left[ g_{f,j} J_1\left(\frac{\Omega_\mu}{\omega_\mu}\right)  \bar{S}_{+,j} \sigma_- +\Lambda_j\xi_j  e^{i\theta_j} \hat{a}^\dag \bar{S}_{-,j} + H.c. \right] \,,\nonumber
\end{split}
\end{equation}
%
where $\dot{\phi}$ is the chirp of the coherent driving. Here we define the detuning $\Delta_j = \omega_{r,j}- \omega_r$ with $\omega_r=\langle \omega_{r,j}\rangle$, $\delta = \omega_L -\omega_c$, $\Delta_0 = \omega_r -\omega_c$ and $\Delta_1 = (\omega_r -\omega_L) - \bar{D}$ with $\bar{D}=\langle D_j\rangle$, $\Delta_q = \omega_q - \delta$. $\langle \cdot \rangle$ means the statistical ensemble average. We have the identity  $\Delta_1 - \Delta_0 = \delta - \bar{D}$. Under the two-photon resonance condition, $\delta = \bar{D}$, and we  also have, $\tilde{\Delta}_q = \Delta_q - \omega_\mu$, $\delta_{en}=\sum_j^N \frac{g_{c,j}^2}{\Delta_0 + \Delta_j}$, $\Delta_{Spin,j}=\left[\Delta_1 - \Delta_0 -\delta_j -\dot{\phi} + \frac{\Omega_c^2}{\Delta_1 + \Delta_j - \delta_j} \right]$, and we define the operators $\bar{S}_{+,j}=|e_j\rangle \langle g_j|$ and $\bar{S}_{-,j} = \bar{S}_{+,j}^\dag$, while $J_1(x)$ is the Bessel function of the first kind. Here we assume that $|\Delta_0|, |\Delta_0| \gg \gamma_j$ with $\gamma_j = \gamma_{0,j} + \gamma_{1,j}$, and set $\Lambda_j = \frac{\Omega_c}{2} \bar{g}_c \left( \frac{1}{\Delta_0 + \Delta_j} +  \frac{1}{\Delta_1 + \Delta_j -\delta_j} \right)$, and $\xi_j =\frac{g_{c,j}}{\bar{g}_c}$, with $\bar{g}_c = \langle g_{c,j}\rangle$. Note that  the detuning $\delta_{en}$ and the parameters $g_{f,j},\xi_j, \theta_j$ are fixed once the setup is fabricated. A major advantage of our sideband transition configuration is that it allows one to modulate the coupling rate within the triparite configuration of flux-spin ensemble-optical mode. This flexibility will later permit us to use  Stimulated Raman Adiabatic Passage (STIRAP) for quantum transfer and this has tremendous advantage as regards robustness to noise and parameter imprecision over  fixed on-resonance coupling schemes for transfer \cite{Zhu:2012js,Zhu:PhysRevLett111p107008}.

We can now consider the swap of quantum information between the flux qubit and the optical cavity mode via the spin ensemble in one interface. 
We first determine a good model for the spin ensemble with inhomogeneous broadening in the transition frequencies and coupling rates.
We divide the spins into $N_g=20$ groups and consider small inhomogeneities between the groups for the coupling rates $g_{c,j}$, and  transitions frequencies $D_j$, and $\omega_{r,j}$. This  model  reproduces quite precisely the Rabi oscillations observed in the experimental observation \cite{Zhu:2012js} (see Supplementary Material), and we use this model for our numerical investigations below. Correspondingly, the Jaynes Cummings coupling rates $g_{f,j}$, and $g_{c,j}$, are the cooperative coupling rates of the $j$th group which is increased by a factor $\sqrt{N_j}$ with respect to the single-spin coupling rate. Unlike the quantum memory by Zhu et al. \cite{Zhu:2012js}, the strength of the overall magnetic coupling rate  between the spin ensemble and the flux qubit is limited by the applicable thickness of the crystal chip hosting the spins. We will find that  crystal chip must be quite thin to avoid degradation of the transfer fidelity due to phase mismatching and this reduced thickness restricts the degree of achievable magnetic coupling.

\begin{figure}
 \centering
\includegraphics[width=0.8\linewidth]{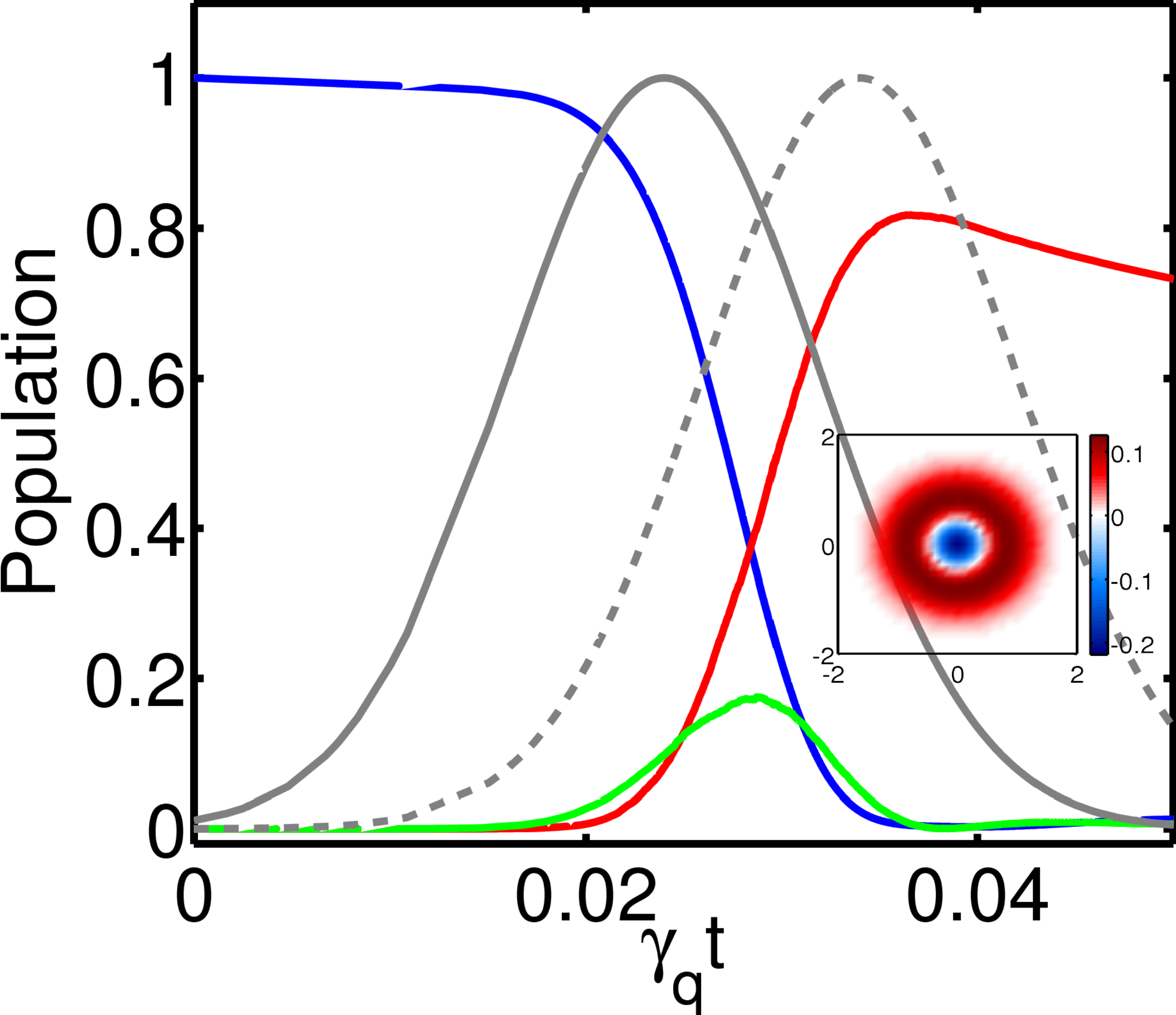} \\
\caption{(Color online) Swap of quantum information from the flux qubit to the cavity mode via the STIRAP protocol.Grey lines are the time modulated STIRAP couplings via a driving classical optical field $\Omega_c(t)$ (solid line) and modulated cavity-QED coupling strength $g_fJ_1[\Omega_\mu(t)/\omega_\mu]$ (dashed line). Blue line shows the population of the excited state of the source flux qubit, red line is the population of the target photonic state $|n=1\rangle$, while green line shows collective excitation of the spins. Insert is the Wigner function of the cavity mode at $\gamma_{1,qb}t=0.036$. Other parameters are $\gamma_{1,qb}=0.4$ \mega\hertz, $\kappa=3$ \mega\hertz~ and $g=105$ \mega\hertz.
}\label{fig:swap}
\end{figure}

To swap quantum information from the flux qubit to the cavity mode we perform a two-photon resonant STIRAP transfer of the population from the flux qubit to the optical cavity, shown in Fig. \ref{fig:swap}. The system is initially populated in the excited state of the flux qubit, $\langle \sigma_{ee}(t=0) \rangle =1$. We modulate the coherent optical driving $\Omega_c$ and the flux bias $\Omega_\mu$ such that two Rabi frequencies are Gaussian functions given by $J_1\left(\Omega_\mu(t)/\omega_\mu\right)=0.58 e^{-(t-\tau_{d,f})^2/2\tau_f^2}$ and $\Omega_c(t)=\Omega_{c,0}e^{-(t-\tau_{d,c})^2/2\tau_c^2}$ with the amplitude $\Omega_{c,0}$, of the coherent classical optical control field. To minimize the operation time, the three subsystems interact on-resonance such that $\dot{\phi} = \Delta_1 - \Delta_0 + \Omega_c^2/\Delta_1 -\delta_{en}$ and $ \dot{\phi} + \tilde{\Delta}_q = -\delta_{en}$ yielding $\omega_\mu = \Delta_q + (\Delta_1 - \Delta_0)+ \Omega_c^2/\Delta_1$. For $\Omega_{c,0}/\kappa ,\langle g_{f,j}\rangle/ \kappa\gg 1$, we can finish the swap operation before the loss of the excitation due to any decay within the system. Here $0.58 \sqrt{N}\langle g_{f,j}\rangle \approx \Lambda_j \approx 105$ \mega\hertz.  
An advantage of this STIRAP transfer is that the excitation of the spins is greatly suppressed.  As a result, the detrimental effect of the inhomogeneous broadening of the spins is very small. 

Figure \ref{fig:swap} shows how well our STIRAP-based SWAP scheme works for $\gamma_{1,qb}=0.4$ \mega\hertz, $\kappa=3$ \mega\hertz.
Such high quality flux qubits \cite{2011NatPh...7..565B,2010PhRvL.105f0503F} and Fabry-P\'erot cavities \cite{HighQFBcavity1} are available using  existing experimental technology. To perform the STIRAP control the pulse $J_1\left(\Omega_\mu(t)/\omega_\mu\right)$ follows the pulse $\Omega_c(t)$ with a delay $\tau_{d,f} - \tau_{d,c} = 1.25\tau_{f}$ and $\tau_{f}=\tau_{c}=3$ \nano\second~ ($0.008\gamma_{1,qb}^{-1}$, see supplementary material). At $\gamma_{1,qb} t=0.036$, the fidelity of the photonic state $|n=1\rangle$ reaches its maximum, $\mathscr{F}=\sqrt{\langle\Psi|\rho|\Psi\rangle}=0.904$ corresponding to a population of $P_{|n\rangle =|1\rangle}=0.817$.
For $\gamma_{1,qb}=0.4$ \mega\hertz, we can achieve a SWAP fidelity larger than $0.81$ if $\kappa \leqslant 5\gamma_{1,qb}$ and $g>40$ \mega\hertz.
A simpler implementation of the STIRAP technology using a constant chip corresponding to the maximal ac Stark shift $\Omega_{c,0}^2/\Delta_1$ can still reaches $\mathscr{F}=0.897$ at $\gamma_{1,qb} t=0.035$.

\begin{figure}
 \centering
\includegraphics[width=0.48\linewidth]{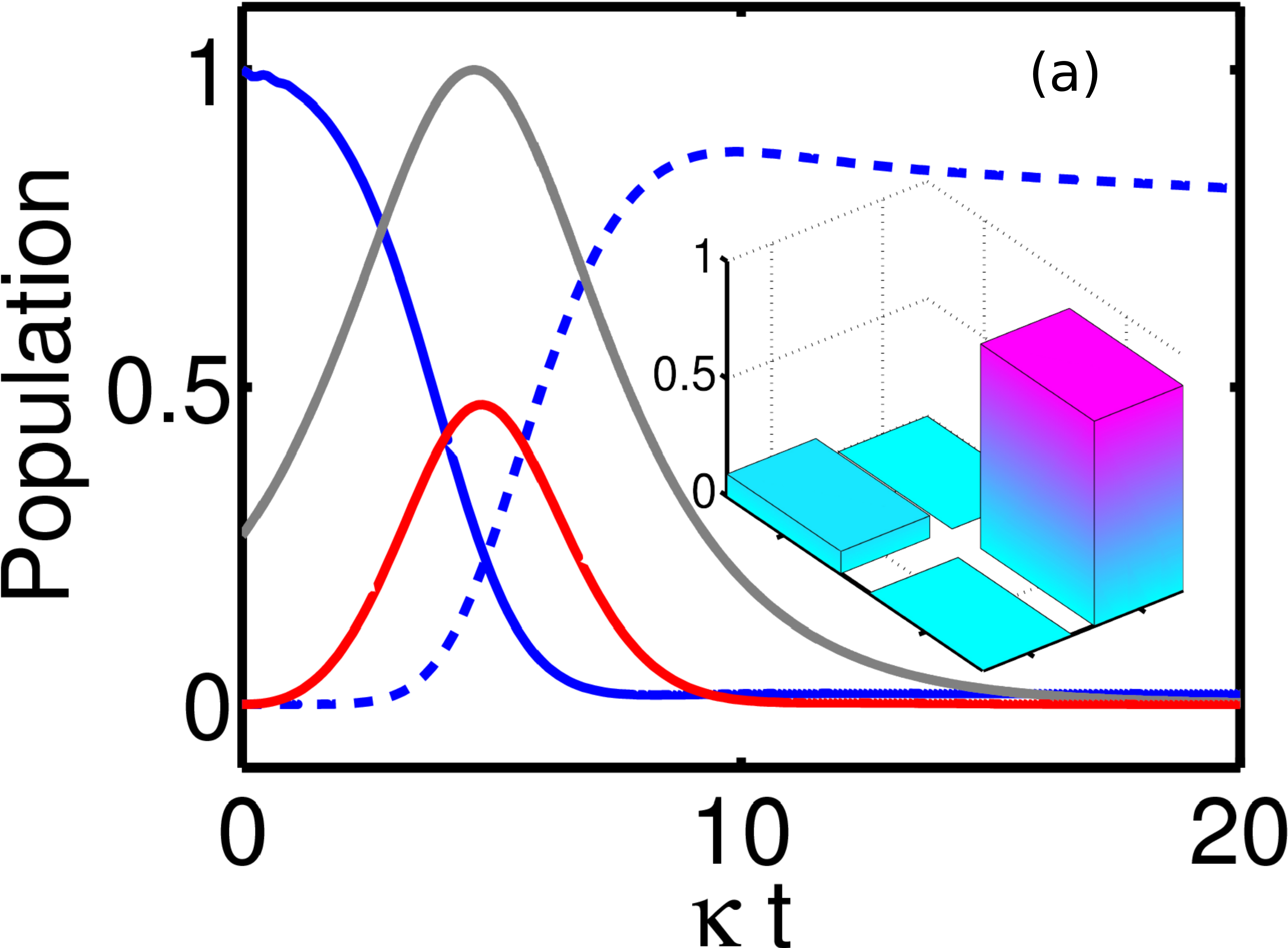} 
\includegraphics[width=0.48\linewidth]{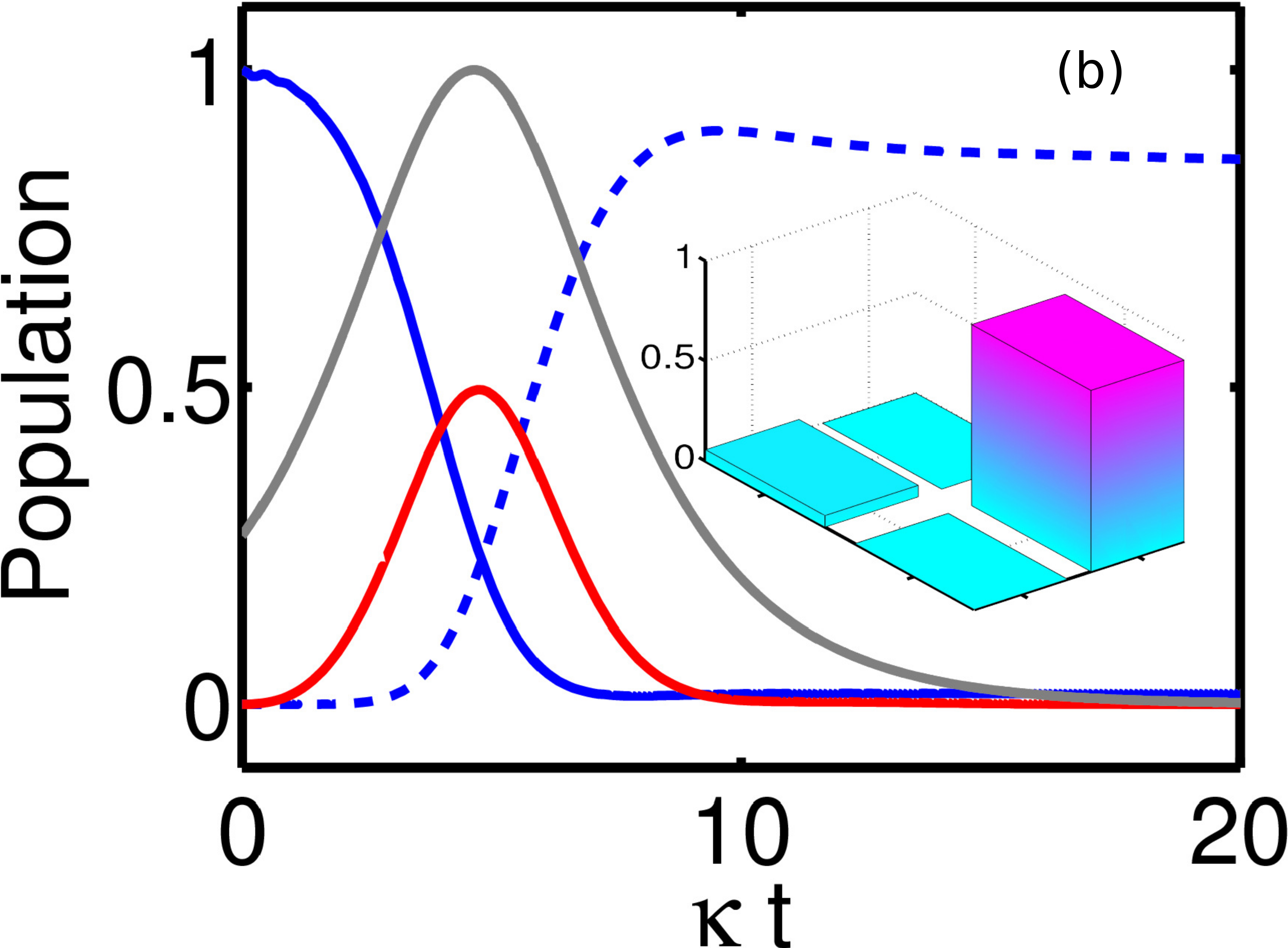} \\
\caption{(Color online) Transfer of quantum information between two distant flux qubits (a) using an ensemble of $\text{NV}^{-}$ centers with $\Gamma_{1,en}=12$ \mega\hertz; while (b) using an ensemble of  $\text{Er}^{3+}$ with $\Gamma_{1,en}\sim 0$. Solid (dashed) blue line graphs the excitation of the source qubit A (target qubit B), red line is the excitation of the antisymmetric state of the combined cavity modes of nodes A and B while the grey line shows the control pulses (identical and simultaneous in both nodes). The inset shows the density matrix of the superconducting qubit at node B at $\kappa t\sim 10$. Other parameters are $\gamma_{1,qb}=20$ \kilo\hertz, $\gamma^*_{2,qb}=0$, $\Gamma^*_{2,en}=0.9$ \mega\hertz, $\sigma_{\Delta}=14.4$ \mega\hertz, $\sigma_\theta=0.1\pi$.}\label{fig:QNet}
\end{figure}

We can finally now seek to use our microwave-optical quantum interface to transfer quantum information between two remote superconducting qubits by exploiting the scheme proposed by Cirac et al. \cite{1997PhRvL..78.3221C},  to construct a quantum internet. To perform the quantum remote transfer we create a tripartite Raman transition where the spin ensemble dispersively couples to the flux qubits and the cavity modes with detuning $20\sqrt{N}g_{f}$. According to the CZKM scheme \cite{1997PhRvL..78.3221C,PRA65p062308}, we need only tune the coupling $\Lambda_j$ by modulating the control field $\Omega_c$ and can set $\Omega_\mu=0$. In the simulation, we eliminate the small time delay in the control field $\Omega_c$ of the node B related to the retardation in the propagation between two nodes \cite{1997PhRvL..78.3221C}. As an example, Fig. \ref{fig:QNet} shows the distant transfer of the quantum state from node A to node B. We assume that two nodes are identical. The flux qubit-spin ensemble coupling rates $\sqrt{N}g_{f,j}= 10 \kappa$ with $\kappa=10$ \mega\hertz~ is constant in time, while we modulate the optical Raman coupling via the classical laser driving field $\Omega_c= \Omega_{c,0}\, {\rm sech}{\left[-(t-\tau_{d,c})/\tau_c\right]}$, with $\Omega_{c,0} = 200\kappa$ and $\Delta_0=20\Omega_{c,0}$. As a result, the excitation in the flux qubit in node A is transferred to the flux qubit in node B with a  fidelity in excess of $90\%$. For $\text{NV}^{-}$ centers, we use the collective decay rate of the ensemble of spins $\Gamma_{1,en}=12$ \mega\hertz~ for fitting the experimental data (see supplementary material). The quantum state of the flux qubit A can be transferred to  qubit B with a fidelity of $\mathscr{F}=93.3\%$ corresponding to a Node B population of $0.87$ at $\kappa t=9.55$. While for the ensemble of $\text{Er}^{3+}$, the spin magnetic excited state decay can be neglected as $T_1=4$ \second \cite{ErMagTran1}, and the inhomogeneous broadening is $13.8$ \mega\hertz. This negligible decay increases the fidelity slightly to $95.1\%$ (Population of $0.904$).

Next we compare the available magnetic coupling rates in two implementations using $\text{NV}^-$ centers or rare earth ions. A more detailed discussion can be found in section V of the supplementary material.
Both the swapping and remote transfer of the quantum information require a strong coupling between the flux qubit and the spins. However the physical size of the spin ensemble is constrained by the area of the flux qubit and also the crystal chip must be quite thin to reduce the detrimental effects of optical phase mis-match on the transfer fidelity. Due to their large electron $g-$factors, rare earth ions like $\text{Er}^{3+}$, can provide a coupling rate of about $400$ \mega\hertz~ which is large enough for our task. However, the largest usable magnetic coupling strength between the flux qubit and the ensemble of $\text{NV}^-$ centers is only about $19$ \mega\hertz. This  coupling rate is low to achieve a high fidelity swap of the flux qubit quantum state. To increase the coupling rate, we can focus the magnetic flux on the small diamond chip using superconducting flux focusing techniques. Flux focusing can greatly increase the magnetic field strength over the focusing region \cite{Ffocuser1,Ffocuser2,Ffocuser3}, and utilising this our scheme can be usefully applied to an ensemble of $\text{NV}^-$ centers to achieve good transfer fidelities.

In summary, we have proposed a theoretical scheme for an all-solid-state quantum interface between microwave and optical quantum information with sufficient tunability to permit local and long-distance high fidelity quantum transfer. We show that by coupling a superconducting circuit magnetically to an ensemble of three level color centers or rare-earth ions, where the latter also couples to an optical cavity  mode, we are able to use the STIRAP technique to swap the quantum state between the superconducting qubit and the optical cavity. Using this quantum interface, we have demonstrated the proof-in-principle quantum network transferring the quantum information with a fidelity larger than $90\%$ between two remote superconducting qubits. This is an essential progress towards the realization of a quantum internet, in particular in the optical telecom C-band around $1550$ \micro\meter~ using the rare earth Er$^{3+}$. 

\section*{Acknowledgements}
This work was partially funded by the Australian Research Council Centre of Excellence for Engineered Quantum Systems (EQuS), project number, CE110001013.


\bibliographystyle{apsrev}

\end{document}